\documentclass[structabstract]{aa}  
\usepackage{graphicx}
\usepackage{natbib}
\usepackage{txfonts}
\begin{document}
\def\frac{$''$\hspace*{-.1cm}}
\def\min{$'$}
\def\deg{$^{\circ}$\hspace*{-.1cm}}
\def\hii{H\,{\sc ii}}
\def\hi{H\,{\sc i}}
\def\hg{H$\gamma$}
\def\hd{H$\delta$}
\def\he{H$\epsilon$}
\def\hb{H$\beta$}
\def\ha{H$\alpha$}

\def\lam{$\lambda$}

\def\aiii{Al~{\sc{iii}}\ }
\def\ariii{[Ar\,{\sc{iii}}]}
\def\ariv{[Ar\,{\sc{iv}}]}
\def\cii{C\,{\sc{ii}}}
\def\ciii{C\,{\sc{iii}}}
\def\civ{C\,{\sc{iv}}}
\def\ci{Ca\,{\sc{ii}}}
\def\cliii{[Cl\,{\sc{iii}}]}
\def\crii{Cr\,{\sc{ii}}}
\def\feii{Fe\,{\sc{ii}}}
\def\feiii{Fe\,{\sc{iii}}}
\def\hei{He\,{\sc{i}}}
\def\heii{He\,{{\sc ii}}}
\def\nii{[N\,{\sc{ii}}]}
\def\niii{N\,{\sc{iii}}}
\def\niv{N\,{\sc{iv}}}
\def\nv{N\,{\sc{v}}}
\def\ni{Na~{\sc{i}}\ }
\def\nei{Ne~{\sc{i}}\ }
\def\neiii{[Ne~{\sc{iii}}]}
\def\oi{[O\,{\sc i}]}
\def\oii{[O\,{\sc ii}]}
\def\oiii{[O\,{\sc iii}]}
\def\sii{[S\,{\sc ii}]}
\def\siii{[S\,{\sc iii}]}
\def\siv{S\,{\sc{iv}}}
\def\si_ii{Si\,{\sc{ii}}}
\def\si_iii{Si\,{\sc{iii}}}
\def\siiv{Si\,{\sc{iv}}}
\def\tiii{Ti\,{\sc{ii}}}

\def\x{$\times$}
\def\av{$A_{V}$}
\def\mcube{$^{-3}$}
\def\cm2{cm$^{-2}$}
\def\sec{s$^{-1}$}

\def\sm{$M_{\odot}$}
\def\slum{$L_{\odot}$}
\def\ab{$\sim$}
\def\sec{s$^{-1}$}

   \title{A very young component in the pre-eminent starburst region of the 
Small Magellanic Cloud}


   \author{M. Heydari-Malayeri
          \inst{1}
          \and
          R. Selier\inst{1}
          }

\institute{Laboratoire d'Etudes du Rayonnement et de la Mati\`ere en Astrophysique (LERMA), \\
Observatoire de Paris, CNRS, 
61 Avenue de l'Observatoire, 75014 Paris, France. \\
              \email{m.heydari@obspm.fr}
             }

\authorrunning{Heydari-Malayeri \& Selier}
\titlerunning{SMC compact \hii\ region N66A}

\date{Received 10 February 2010 / accepted 14 April 2010}

 
   \abstract
   {Despite extensive research on various components of the N66/NGC\,346 complex, 
few studies have so far focused on N66A, which is a special object in 
the whole complex and therefore deserves scrutiny. The study of this compact \hii\ 
region and its fellow objects seems important 
in the framework of massive star formation in the Magellanic Clouds.}
   {We present a study of the compact \hii\ region N66A in the SMC pre-eminent
starburst region N66/NGC\,346. 
}
   {This analysis is based mainly on our optical ESO NTT observations,  
both imaging and spectroscopy, 
coupled with archive  {\it HST} {\rm ACS data and Spitzer infrared images 
(IRAC 3.6, 4.5, 5.8, and 8.0\,$\mu$m).} 
}
   {We derive a number of physical characteristics of the compact \hii\ region N66A.
For the first time using spectroscopy we present the spectral classification of 
the main exciting star of N66A. 
Its spectral features indicate that it is a main-sequence massive star of type O8. 
We compare this result 
with that based on the stellar Lyman continuum flux estimated from 
the ionized gas \hb\ flux.  
The compact \hii\ region belongs to a rare class of \hii\ regions in the 
Magellanic Clouds, called high-excitation blobs (HEBs). We propose that 
N66A probably represents a very young massive star 
formation event in the N66 complex, that has a range of ages.   
 }
   {}


   \maketitle
%

\section{Introduction}

LHA\,115-N66, or in short N66 \citep[][]{Henize56}, is the largest and the most 
luminous \hii\ region in the Small Magellanic Cloud (SMC).  It is also
known as DEM S\,103 (Davies et al. 1976) and NGC\,346, the latter 
referring to the bright OB association located at its center. 
We refer the reader to the magnificent images of the region taken with the Advanced Camera for 
Surveys (ACS) onboard the {\it Hubble Space Telescope} and the {\it JHK}  
composite obtained with ISAAC on the ESO VLT telescope   
\citep[][]{Nota06,Gouliermis09}. 
N66 is considered to be the scaled-down counterpart of the Large 
Magellanic Cloud starburst 30 Doradus. It indeed hosts the largest sample of 
young, massive stars in the whole SMC with 33 O-type stars among which 11 are of 
type O6.5 or earlier \citep[][]{Massey89,Walborn00,Evans06}. 
It contains at least one W-R star in the massive binary or maybe triple system HD\,5980.  
An age of \ab\,3 Myr has been estimated for NGC\,346 from evolutionary models in 
the H-R diagram \citep[][]{Bouret03}. HD\,5980 lies behind a SNR 
\citep[B0057-724,][]{Reid06,Naze02,Danforth03,Naze04}, which has no known optical counterpart. 
Compared to the Orion Nebula, N66 has 
an \ha\ luminosity almost 60 times higher \citep[][]{Kennicutt84}. This radiant flux is 
also reminiscent of those of giant \hii\ regions in distant metal-poor galaxies, such as 
regions A1 and A2 in IC 4662 lying 2.44 Mpc away 
\citep[][and references therein]{MHM90,Crowther09}. 
Therefore, N66 offers a valuable template for studying these 
kinds of distant galaxies with high resolution.  

Apart from its recently formed massive star population mentioned above, 
NGC\,346 also has a large population of low-mass, pre-main-sequence stars 
\citep[][]{Gouliermis06,Nota06,Sabbi07,Hennekemper08} covering a mass range down to the 
subsolar regime. 
The PMS population is found to be mainly concentrated in a number of subclusters
away from the massive star association \citep[][]{Sabbi07,Schmeja09}.
The typical ages of the PMS population derived from models appear to suggest  
that low-mass star formation events occurred at two different epochs 
about 4 and 10 Myr ago \citep[][]{Hennekemper08}.

Being such an extraordinary object, the N66 complex has been extensively studied 
from various viewpoints using almost the whole electromagnetic range,  
from radio wavelengths to X-rays. More specifically, the studies have used       
the radio continuum \citep[][]{Ye91,Reid06} 
and the 21 cm atomic line  \citep[][]{Staveley-Smith97,Stanimirovic99}, 
the CO line with the SEST telescope \citep{Rubio96,Rubio00}, 
the infrared with both the Infrared Space Observatory 
\citep{Contursi00} and the Spitzer Space Telescope 
\citep{Bolatto07, Simon07},  optical spectroscopy 
\citep[][]{Massey89,Walborn00,Evans06},   
ultraviolet and optical spectra \citep{Walborn00}, 
with data from the {\it HST} spectrograph STIS, AAT, and ESO 3.6 m telescope,  
imaging with ACS and WFC on board {\it HST} \citep[][]{Nota06}, \ha\ 
\citep[e.g.,][]{Kennicutt84,LeCoarer93,Smith00}, in the UV with the 
International Ultraviolet 
Explorer \citep[][]{deBoer80} and the Far-Ultraviolet Spectroscopic Explorer 
 \citep[][]{Danforth03}, and in the X-rays with XMM-Newton and Chandra 
\citep[][]{Naze02,Naze04}. 

The present study is concerned with massive star formation in the N66 complex. 
Clustered mainly in NGC\,346, as mentioned above, massive stars dominate the central 
part of the whole \hii\ region with their strong UV radiation field. Twenty-two 
of the above-mentioned 33 O stars are contained in the central cluster. The hottest 
star, W3, is reclassified as O2 III(f*) \citep[][]{Walborn02a,Walborn02b}.
The most massive star, W1, of the central cluster, classified 
O4 III(n)(f)   \citep[][]{Walborn86},  
has multiple components and the mass of the brightest component is at most 85 \sm\, 
 \citep[][]{MHM91}.  
 The cluster has disrupted the bulk of the natal molecular cloud, and therefore
not much CO emission is detected towards N66, except for two positions which are   
mapped in the (1-0) and (2-1) transitions \citep{Rubio96,Rubio00}.

One of these CO peaks is associated with a remarkable feature of the whole 
landscape, a compact \hii\ region, called N66A, according to \citet[][]{Henize56}.   
The \hii\ region apparently lies at the south-eastern end of 
an absorption lane that runs over some 60 pc from north-west to south-east below 
the NGC\,346 cluster. This paper is mainly devoted to this compact \hii\ region.  
Despite extensive research on various components of the N66/NGC\,346 complex, 
few studies have so far dealt with this \hii\ region. We attempt to demonstrate that 
this region represents the youngest episode of massive star formation in N66.  

A word of caution seems necessary about the name of this object.  
From their observations of H$_{2}$ emission line (2.14 $\mu$m) and the 
ISOCAM LW2 band (centered at 6.75 $\mu$m),   
\cite{Contursi00} and \cite{Rubio00} detected several embedded sources towards N66, which 
they alphabetically designated from ``A'' to ``I''. 
The  IR source A should not be confused with 
the Henize N66A \hii\ component, which corresponds to the IR source ``H''. 
In a similar way, NGC\,346 corresponds to ``C''.

The paper is organized as follows. Section 2 presents the observations, 
data reduction, and the archive data used ({\it HST} ACS data, Spitzer data). 
Section 3 describes our results (morphology, physical parameters, spectral classification). 
Section 4 presents our discussion, and finally our conclusions are summarized in Sect. 5.

\section{Observations and data reduction}

\subsection{NTT Imaging}

N66 was observed on 28 September 2002 using the
ESO New Technology Telescope (NTT) equipped with the active optics and
the SUperb Seeing Imager (SuSI2). The detector consisted of two CCD
chips, identified as ESO \#45 and \#46.  The two resulting frames were 
automatically combined in a single FITS file, whereas the space
between the two chips was ``filled'' with some overscan columns so that
the respective geometry of the two chips was approximately
preserved. The gap between the chips corresponds to \ab\,100 true CCD
pixels, or \ab\,8\frac.  The file format was 4288\,\x\,4096
pixels, and the measured pixel size  0\frac.085 on the sky. Each
chip of the mosaic covered a field of 5\min.5\,\x\, 2\min.7. We refer to the
ESO manual SuSI2 for more technical information 
(LSO-MAN-ESO-40100-0002/1.9).

\begin{figure*}[]
  \centering
\includegraphics[width=1.0\hsize]{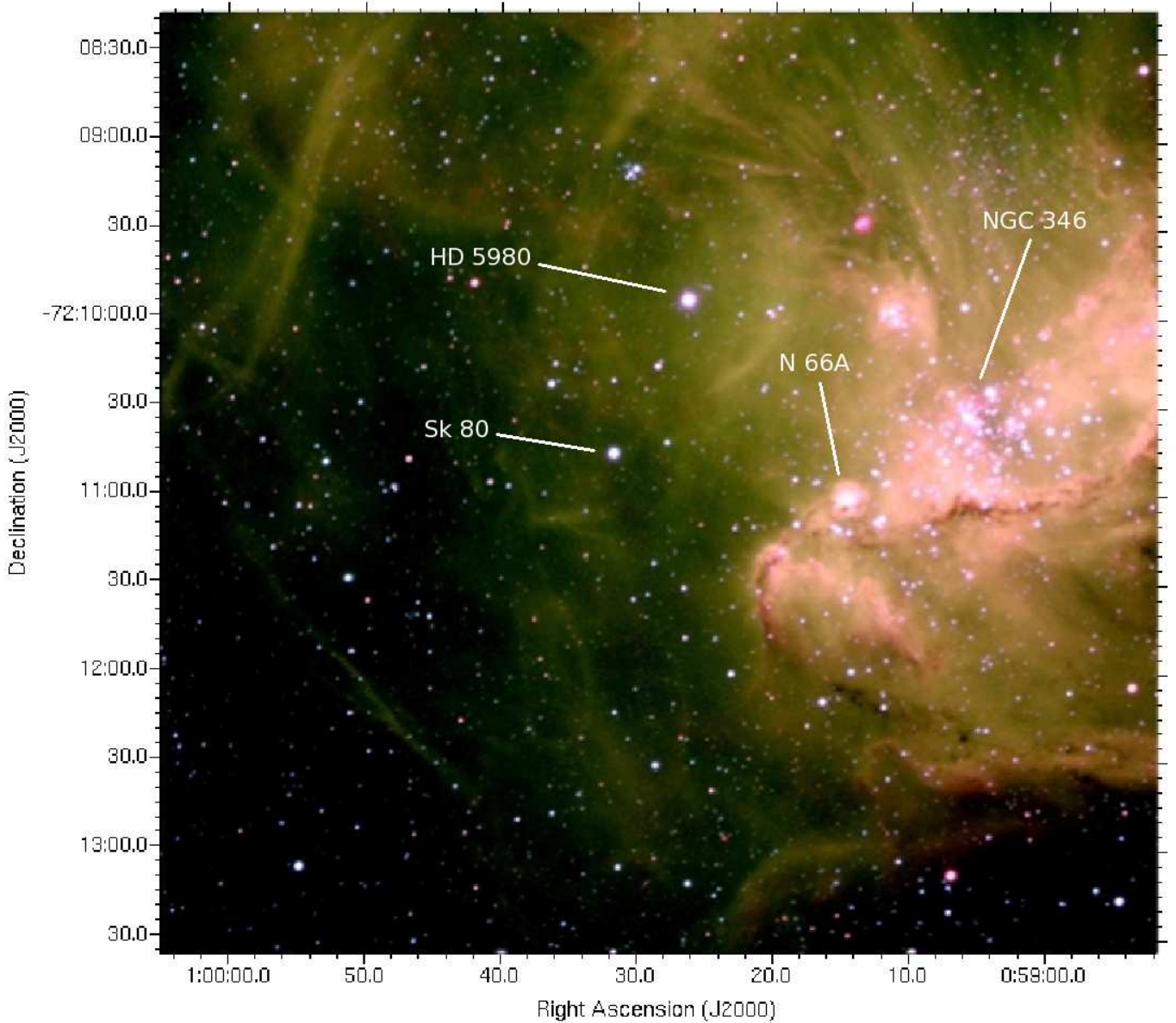}
\caption{A composite three-color image of the SMC \hii\ region N66. 
The star cluster above the curling absorption lane is the OB association NGC\,346. N66A is the 
brightest compact \hii\ region lying at the eastern end of the dark lane.  
Note the wind-driven bubble centered on the brightest star HD\,5980. 
The other bright star lying towards the field center is Sk\,80.
The  image, taken with the ESO NTT/SuSI2, results from the coaddition of narrow-band 
filters \ha\ (red),  \oiii\
(green), and \hb\ (blue). 
The field size is 336\frac\,\,\x\,350\frac\, corresponding to 
100\,\x\,103 pc. North is up and east to the left. } 
\label{fig:ntt}
\end{figure*}

Nebular imaging was carried out using the narrow-band filters centered
on the emission lines \ha\, (ESO \#884), \hb\, (\#881), and
\oiii\,(\#882) with three basic exposures of 300 sec. 
The image quality was quite good during the night, being represented 
by a seeing of 0\frac.5--0\frac.8 (Fig.\,\ref{fig:ntt}).   
We constructed the line ratio maps \ha/\hb\, and  \oiii/\hb\  
from nebular imaging. We also took two exposures using filters ESO
\#811 ($B$), \#812 ($V$), and \#813 ($R$) with unit 
exposure times of 15 sec for $B$ and $V$ and 10 sec for $R$,  
respectively. The exposures for each filter were repeated two times 
using ditherings of 5\frac\,--10\frac\, for bad pixel
rejection.

\subsection{NTT spectroscopy}

The EMMI spectrograph attached to the ESO NTT telescope was used on 29
September 2002 to obtain several long-slit spectra of the
stars.  The grating was \#\,12 centered on 4350\,\AA\, (BLMRD mode)
and the detector was a Tektronix CCD TK1034 with 1024$^{2}$ pixels of
size 24 $\mu$m.  The covered wavelength range was 3810-4740\,\AA\, 
and the dispersion 38\,\AA\,mm$^{-1}$, giving {\sc fwhm} 
resolutions of $2.70\pm0.10$  
pixels or $2.48\pm0.13$\,\AA\, for a 1\frac.0 slit. 
At each position, we took three 10 min exposure. 
The instrument response was derived from 
observations of the calibration stars  LTT\,7379, LTT\,6248, and  LTT\,7987. 
The seeing conditions varied  between 0\frac.9 and 1\frac.2.
The identifications of the stars along the
slits were based on monitor sketches drawn during the observations.

Furthermore, EMMI was used on 26 and 27 September 2002 to obtain nebular 
spectra with gratings \#\,8 (4550-6650\,\AA) and \#\,13 4200-8000) in the REMD mode 
and with grating \#\,4 (3650-5350\,\AA) in the BLMD mode. In the REMD mode,    
the detector was  CCD \#\,63, 
MIT/LL, 2048\,\x\,4096 pixels of  15$^{2}\,\mu$m$^{2}$ each.   
Spectra were obtained with the slit set in 
east-west and north-south orientations using a basic exposure time 
of 300 sec repeated several times. The seeing conditions varied between 
0\frac.6 and 0\frac.8. 
Reduction and extraction of spectra were performed using the IRAF software
package. Fluxes were derived from the extracted spectra with the 
IRAF task SPLOT. The line fluxes were measured by fitting 
Gaussian profiles to the lines as well as by simple pixel integration in 
some cases. The nebular line intensities were corrected for interstellar 
reddening using the formulae given  
by \citet{Howarth83} for the LMC extinction.

\subsection{Archive data}

\subsubsection{{\it HST} ACS data}

We used archive imaging data of NGC\,346 (GO\,10248, P.I. A. Nota). 
These observations were obtained with ACS (Advanced Camera for 
Surveys) onboard {\it HST}. The images were taken with the 
Wide Field Camera (WFC) using broad- and
narrow-band filters (F555W, F814W, F656N) in 2004 July. They were 
used to examine the morphology of the compact \hii\ region N66A and 
in particular resolve its exciting stars. In addition, we 
produced a composite image of N66A (Figs.\,\ref{fig:HST_neb}, \ref{fig:HST_stars}) 
and also used the photometry derived from these observations (Gouliermis 2006).
We also used the archive observations obtained in July 2006 using 
ACS with the High Resolution Channel (HRC) and the ultraviolet filters F220W and F330W 
(GO\,10542, P.I. A. Nota).  

\subsubsection{Spitzer data}

The Spitzer archive data used in this paper come from the S$^{3}$MC
project. This is a project to map the star-forming body of the SMC
with Spitzer in all seven Infrared Array Camera (IRAC) and Multiband
Imaging Photometer for Spitzer (MIPS) bands. The MIPS data were
obtained in 2004 November and the IRAC data in 2005 May. We used
the IRAC data to build a composite image of N66 
and also obtain the photometry of N66A. 
The typical PSF of the IRAC images in the 3.6, 4.5, 5.8, and 8.0 $\mu$m bands 
is 1\frac.66 to  1\frac.98 and that of  MIPS at 24 $\mu$m is 6\frac\, 
\citep[][]{Bolatto07}. The derived photometry for N66A in the 3.6, 4.5, 5.8, and 8.0 $\mu$m 
bands are 12.22, 11.82, 10.25, and 8.66 mag, respectively, using an integration 
aperture of 5 pixels, or 6\frac\, in radius. Measurements with either slightly larger 
or smaller apertures do not affect the color results.  
Moreover, we tried to detect substructures in the Spitzer 
images of N66A (see Sect. 4).

\section{Results}

\subsection{Morphology}

Figure 1 displays a composite image of the N66 region taken with the NTT telescope (see 
Sect. 2.1). The field is \ab\,5\min\,\,\x\,5\min\, corresponding to \ab\,90 pc$^{2}$ 
for a distance of about 60 kpc, or  {\it m\,-\,M} = 18.94 mag
\citep[][]{Laney94}. Although the resolution is less than that of the {\it HST} ACS 
image (Sect. 2.3.1), almost all the features of the \hii\ complex are visible. 
The NGC\,346 cluster appears to be at the center of an \hii\ bowl, the southern 
border of which is delineated by a compressed ionized gas front and an absorption 
lane running over some 60 pc. 
In particular, N66A stands out as the most compact \hii\ nebula of the whole region, 
with coordinates (J2000.0) $\alpha$ = 00:59:14.8, $\delta$ = -72:11:01.  
The compact \hii\ region is apparently associated with the compressed gas front and the 
absorption lane.  

The field of view of the NTT image is larger than that of  {\it HST} ACS. 
It also displays a turbulent environment 
in the eastern side of N66 with many indications of shocked gas. In particular,   
the wind-driven bubble centered on HD\,5980 is quite impressive.  
A narrow ridge can also be discerned towards the southern 
outer boundary of the complex. This feature is also affected by stellar shock winds, 
as indicated by its remarkable \sii\ emission \citep[][]{Reid06}. 
Observations with the Australia Telescope Compact Array (ATCA) 
and Parkes Observatory at the Australia Telescope National Facility (ATNF)
also detect an \hi\ cloud concentrated towards this part of N66  
\citep[][]{Staveley-Smith97,Stanimirovic99,Gouliermis08}. 
The \hi\ cloud  is probably in contact with the ridge.

Figure \ref{fig:HST_neb} presents a high-resolution composite image of the N66A 
\hii\ region extracted from the above-mentioned {\it HST} ACS archive data.
The compact \hii\ region is \ab\,10\frac\, in diameter, corresponding to \ab\,3 pc. 
It contains a strong absorption lane.  Interestingly, two bright stars, labelled 
\#1 and \#2, are located towards the central part of the region, above the dust lane 
(see also Fig.\,\ref{fig:HST_stars}). Separated by 0\frac.7 (\ab\,0.2 pc),    
they are the main exciting stars of the \hii\ region, as shown in Sect. 3.3.   
A number of fainter stars are seen across the face of the region. 
These stars are quite bright on the {\it HST} ACS/HRC image in the ultraviolet 
obtained with F220W and F330W filters.  
Five other stars in the UV images (\#3 to \#7) should also be OB stars associated with 
N66A, as suggested by their photometry,  presented in Table \ref{table:A},  which 
also displays the positions.

\subsection{Physical parameters}

The total H$\beta$ flux of N66A was derived using the following 
procedure. First we calculated the relative H$\beta$ flux in an 
imaginary 1\frac\, slit passing through the H$\beta$ image with respect to 
the total flux emitted by the whole N66A region. This value was then 
compared with the absolute flux obtained from the spectra. In both 
cases, a mean flux measured for the NS and EW orientations of the slit 
was used. The total H$\beta$ flux thus obtained was
$F$(\hb )\,=\,2.0\,\x\,10$^{-12}$ erg cm$^{-2}$ s$^{-1}$. 
Considering the extinction law for the LMC  \citep{Howarth83}, we computed the 
reddening corrected intensity $I$(\hb )\,=\,3.6\,\x\,10$^{-12}$ erg cm$^{-2}$ s$^{-1}$. 
We derived a luminosity of 1.6\,\x\,10$^{36}$ erg s$^{-1}$, or 400 \slum, for 
N66A at H($\beta$). 
This luminosity corresponds to a flux of 3.9\,\x\,10$^{47}$ \hb\  photons s$^{-1}$, or  
a Lyman continuum flux of 3.4\,\x\,10$^{48}$ 
photons s$^{-1}$ for the star, assuming that the \hii\ region is ionization bounded. 
The exciting star needed to provide this flux is of spectral type 
about O7.5\,V  \citep[][]{Martins05} (see also Sect. 3.3). 

A number of the derived physical parameters of N66A are summarized in 
Table\,\ref{tab:param}. The mean angular radius of the \hii\ region, corresponding to 
the FWHM of cross-cuts through the \ha\ image, is given in Col. 1. The corresponding 
linear size, obtained using a distance modulus of 
{\it m\,-\,M} = 18.94 mag \citep[][]{Laney94} is presented in Col. 2.The dereddened 
\hb\ flux obtained from a reddening coefficient {\it c}(\hb\,) = 0.26 is given in Col. 3. 
The electron temperature calculated from the forbidden lines ratio 
\oiii\,\lam \lam\,4363/(4959 + 5007), with an accuracy of 4\%, is given in Col. 4. 
The electron density, estimated from the ratio of the \sii\ doublet  \lam \lam\,6717/6731, 
 is presented in Col. 5. It is accurate to \ab\,80\%.  
It is well-known that the  \sii\ lines characterize the low-density 
peripheral zones of \hii\ regions (see below for corroboration). 
Column 6 gives the rms electron density calculated from the \hb\ flux, the radius, $r$, and 
the electron temperature, {\it T$_{e}$}, assuming that the \hii\ region is 
an ionization-bounded Str\"omgren sphere. 
For comparison reasons, an electron density of 50\,$\pm$\,30 cm$^{-3}$ was found from 
our long-slit spectra far from N66A, towards the inner parts of the N66 giant region. 
This agrees well with the mean value of 60 cm$^{-3}$, as estimated 
for the whole N66 region by  \citet[][]{Tsamis03} using the above-mentioned 
ratio of the sulphur doublet, as well as a density of 80  cm$^{-3}$ for N66 reported 
by \citet[][]{Reid06}.  Using the line ratio 
\cliii\,\lam \lam 5537/5517, which identifies denser zones, a density of 
3700 cm$^{-3}$ is found for N66 \citep[][]{Tsamis03}.
Furthermore,  
the total mass of the ionized gas, calculated from the {\it $<$n$_{e}$$>$} 
with the previously noted Str\"omgren sphere assumption is presented in Col. 7. The ionization 
is produced by Lyman continuum photon flux given in Col. 8.  

The most extincted part of the N66A region is the dust lane (Sect.  3.1), where 
the \ha\,/\,\hb\ ratio reaches a value of 5.0 corresponding to A$_{v}$ = 1.7 mag. Outside the
lane, the ratio is smaller in particular around the main exciting stars. The average value of
the Balmer decrement towards N66A is about 3.7 (A$_{v}$ = 0.8 mag). We  note that this ratio
is higher than towards other areas of the \hii\ region N66.
As for the \oiii\,/\,\hb\ ratio, it fluctuates around 4.5 and reaches at the north-eastern 
border of N66A the value of 5.

\subsection{Spectral classification}

Spectral classification was derived for five massive stars 
towards N66A. These stars were extracted from long-slit spectra.  
Four spectra belong to grating \#12 (N66A-1, MPG 455, MPG 595,  
MPG 655) and one spectrum (MPG 445) to
grating \#4. The identification of stars along the slits were based on
monitor sketches drawn during the observations. 
The spectral classification was performed using the criteria stated by \citet[][]{Walborn90}. 
The rectified spectrograms, corrected for the nebulosity background,  
are displayed in  Fig. \ref{fig:sp}. The results are
summarized in Table \ref{table:sp}, which also gives the corresponding astrometric 
and photometric information as well as classifications from previous studies when 
available.  We note that the broader lines of MPG\,445  
is an instrumental effect (grating \#4). 

A particularly interesting star in this study, i.e. N66A-1, was classified as 
O8\,V.  This spectral type agrees well with that based on the stellar Lyman continuum 
derived from the \hb\, emission of the \hii\ region. However, the latter result is 
probably an underestimate because some of the ionizing photons were missed. 
In particular, local dust absorbs a portion of the photons and, in addition, 
since the \hii\ region is not fully ionization-bounded, some of the Lyman continuum photons 
escape into the interstellar medium. This implies that the spectral classification 
based on spectroscopy is also an underestimate. This is probably because 
the spatial resolution of the spectra does not match that of the {\it HST} ACS 
images, which show that the main exciting stars are separated by 0\frac.7. In other words, 
the spectrum of N66A-1 is contaminated by N66-2. If star \#2 were of later type than 
star \#1, the latter would be earlier than O8\,V.

\begin{table*}
\caption{Some physical parameters of N66A}
\label{tab:param}
\begin{tabular}{c c c c c c c c} 
\hline\hline
$\theta$ & $r$  & $I$(\hb )$^{\dag}$        & {\it Te}    & {\it Ne}   & {\it $<$n$_{e}$$>$}  
& {\it M$_{gas}$}  & $N_{L}$ \\
(\frac\,) & (pc) & erg s$^{-1}$ cm$^{-2}$   & (K) & cm$^{-3}$ & cm$^{-3}$ & (\sm ) & ph s$^{-1}$ \\
         &      & \x\,10$^{-12}$ &      &      &      &        &  \x\,10$^{48}$ \\
\hline
    5    &    1.5    &     3.6         &    12770     &  220    &  200    &    90      &  3.4 \\
\hline
\end{tabular} \\

$\dag$ Using a reddening coefficient {\it c}(\hb\,) = 0.26.
\end{table*}

\begin{table*}
\caption{Positions and photometry of stars associated with N66A }             
\label{table:A}      
\begin{tabular}{c c c c c c c c r }     
\hline\hline
Star ID$^{\dag}$ & $\alpha$ (2000) & $\delta$ (J2000) & F220W & F330W & $V$ & $I$ & 
(F220W - $V$) & Other designations$^{\ddag}$ \\
\hline
N66A-1 & 00:59:14.93 & -72:11:01.90 & 13.89 & 14.63 & 15.97 & 16.15 & -2.08 & G 67; MPG 646 \\
N66A-2 & 00:59:14.77 & -72:11:01.97 & 15.64 & 16.34 & 17.33 & 17.41 & -1.69 & G 271; MPG 641 \\
N66A-3 & 00:59:15.38 & -72:11:01.25 & 16.26 & 16.86 & 17.77 & 17.94 & -1.51 & G 427; MPG 654 \\
N66A-4 & 00:59:15.10 & -72:10:59.02 & 16.68 & 17.15 & 17.82 & 17.87 & -1.14 & G 414; MPG 648 \\
N66A-5 & 00:59:15.25 & -72:11:00.78 & 18.56 & 18.59 & 18.74 & 18.45 & -0.18 & G 844; MPG 648 \\
N66A-6 & 00:59:13.74 & -72:10:59.23 & 18.22 & 18.54 & 18.98 & 19.05 & -0.76 & G 1480; MPG 629\\
N66A-7 & 00:59:14.61 & -72:10:59.63 & 19.14 & 19.26 & 19.81 & 19.83 & -0.67 & G 2486; MPG 640\\
\hline   
\end{tabular} \\

$\dag$ See Fig. \ref{fig:HST_stars} for star positions \\
$\ddag$
G: stands for \citet[][]{Gouliermis06}; 
MPG: refers to \citet[][]{Massey89} 
\end{table*}

\begin{table*}
\caption{Stars classified towards N66A}             
\label{table:sp}      
\begin{tabular}{l l l l l l l l}
\hline\hline
Star ID & $\alpha$ (2000) & $\delta$ (J2000) & $V$ & {\it V\,-\,I} &  {\it B\,-\,V} &
Previous classification$^{\dag}$ & This work \\
\hline
N66A-1  & 00:59:14.93 & -72:11:01.9 & 15.97 & -0.18 &  &  & O8 V    \\
MPG 445 & 00:59:04.79 & -72:11:03.1 & 15.33 & & -0.18 & O7.5 V (a) & O8 V     \\
MPG 455 & 00:59:05.44 & -72:10:42.4 & 15.25 & & -0.19 &  & O9.5 V   \\
MPG 595 & 00:59:11.91 & -72:10:55.8 & 15.62 & & -0.20 & B0 V (b)  & B0-0.5 V  \\
MPG 655 & 00:59:15.51 & -72:11:11.6 & 14.82 & & -0.15 & O6 V (b) & O5 V + OB  \\
\hline \\      
\end{tabular}\\
$\dag$ (a):  \citet[][]{Niemela86}; b:  \citet[][]{Massey89} 
\end{table*}

\begin{figure*}[]
  \centering
\includegraphics[width=10cm]{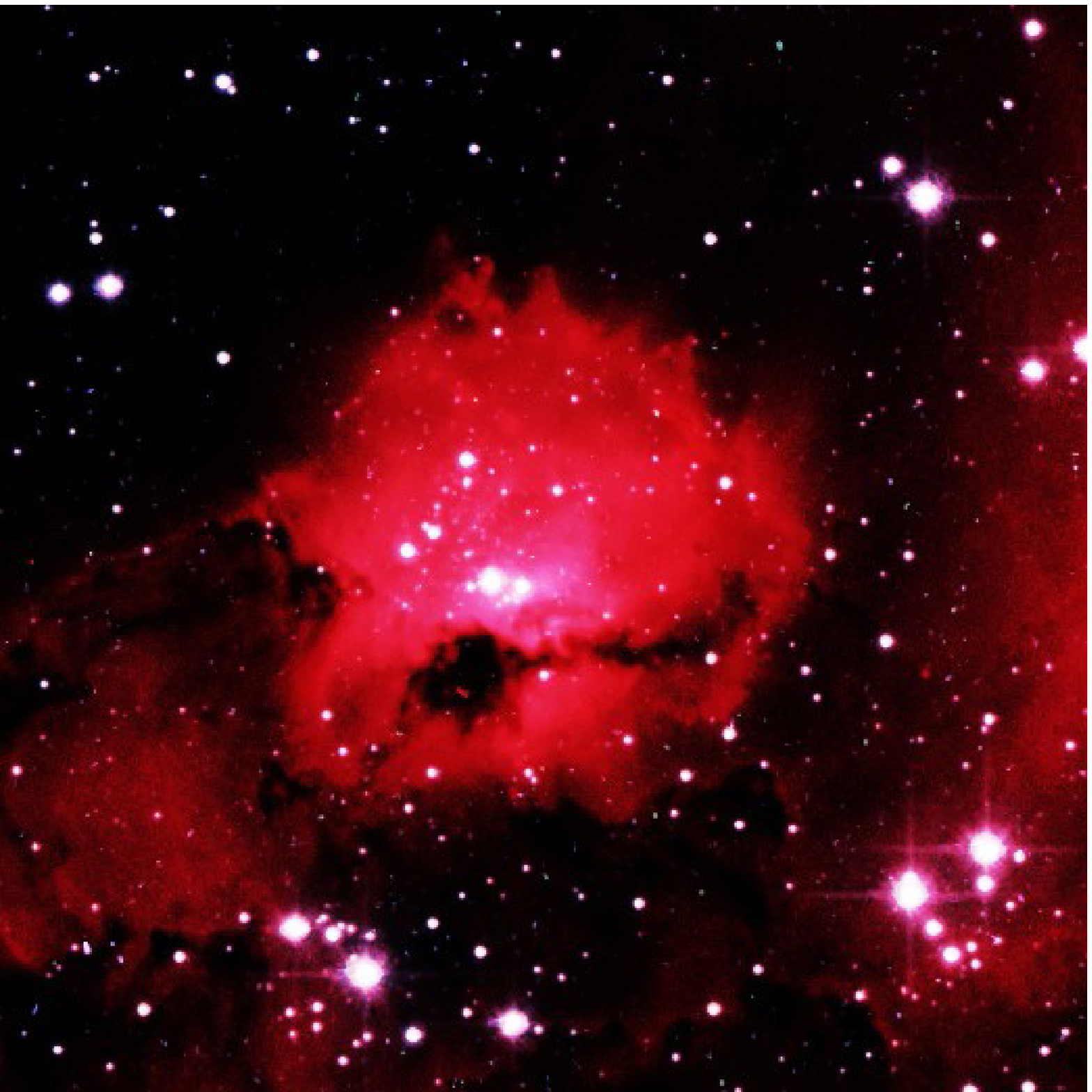}
  \caption{A composite three-color image of SMC N66A created using the {\it HST} ACS 
images in \ha\ (red), filter {\it I}, F814W (green), and filter {\it V}, F555W (blue).  
Field size 521\,\x\,521 pixels, or 26\frac\ x 26\frac\, (\ab\ 7.5\,\x\,7.5 pc). 
North is up and east to the left. }
\label{fig:HST_neb}

\centering
\includegraphics[width=10cm]{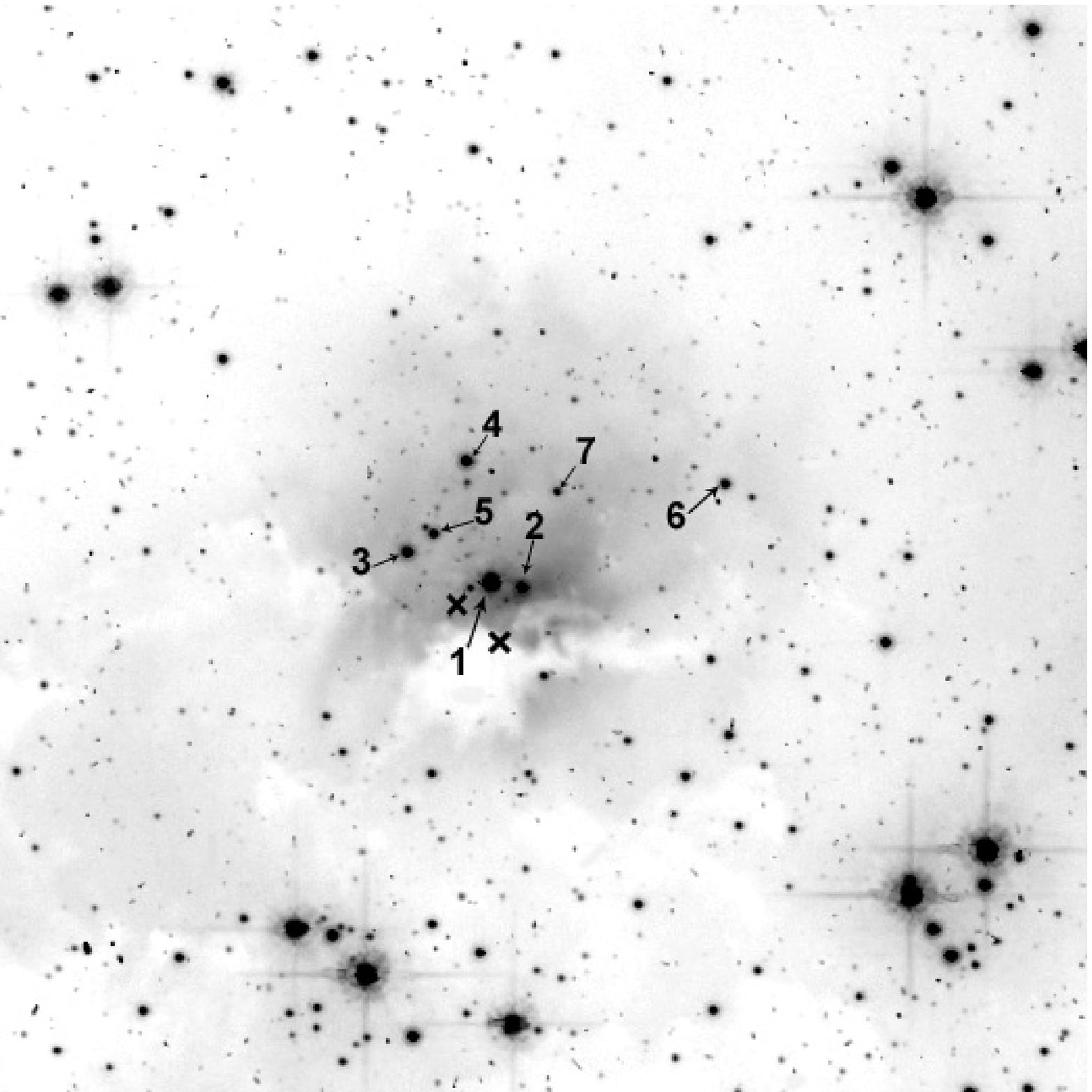}
\caption{The N66A image of SMC N66A extracted from the {\it HST} ACS image obtained 
through filter {\it V} (F555W). The main stars associated with the \hii\ region are 
marked by numbers; see also Table \ref{table:A}. The crosses represent the two YSO positions as 
suggested by Spitzer 
observations \citep[][]{Simon07}. Field size and orientation same as in Fig.\,\ref{fig:HST_neb}. } 
\label{fig:HST_stars}
\end{figure*}

\begin{figure*}[]
\centering
\includegraphics[width=1.0\hsize]{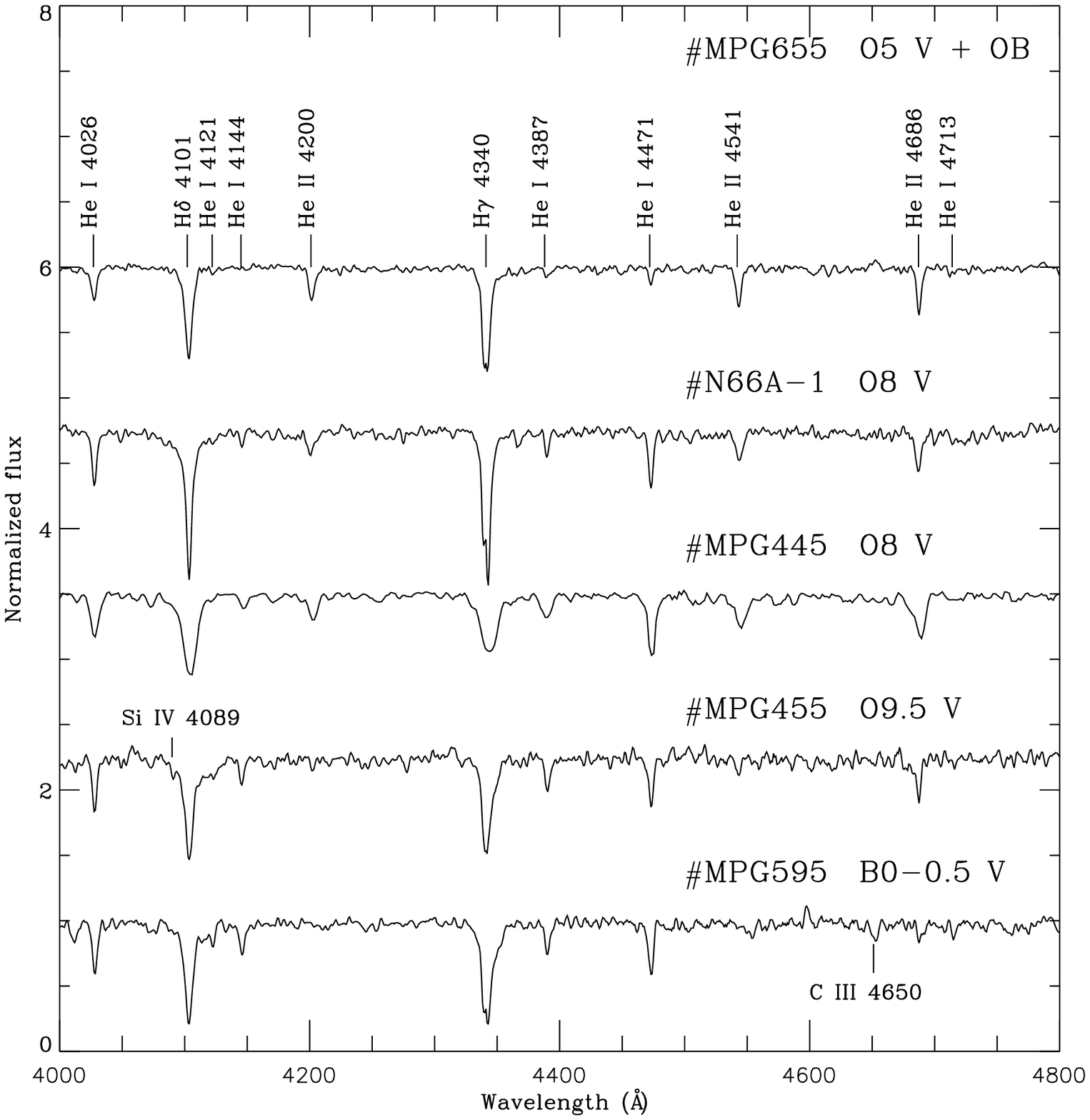}
\caption{Spectrograms of the classified massive stars towards SMC N66A.  }
\label{fig:sp}
\end{figure*}

\section{Discussion}

N66A is clearly the most compact \hii\ region of the N66 complex in 
the optical.  Its relative compactness, brightness, and location 
suggest that it is probably a relatively younger generation in the N66 
complex. It should belong to a distinct and rare class of \hii\ 
regions in the Magellanic Clouds (MCs) called High-Excitation 
``Blobs'', or HEBs \citep[see][for a review]{MHM10}.
In contrast to the typical \hii\ regions of 
the MCs, which are extended structures with sizes of several arc 
minutes corresponding to physical scales of more than 50\,pc and 
powered by a large number of exciting stars, HEBs are relatively dense and small 
regions of $\sim$\,5\frac\, to 10\frac\, in diameter in the optical, 
corresponding to $\sim$\,1.5 to 3.0\,pc  and excited by a much smaller number 
of massive stars. Their excitation, derived from the 
\oiii\,(\lam\,\lam\,4959\,+\,5007)\,/\,\hb\ ratio, is generally 
larger than that of ordinary MC \hii\ regions. 
For a fixed metallicity, the \oiii\,/\,\hb\ ratio increases with the effective 
temperature of the exciting star, 
as well as with the gas density in homogeneous photoionization models. 
      
These compact \hii\ regions are also heavily affected by local dust compared to other 
ionized features of the complex in which they are hosted 
\citep[][]{MHM88,Israel91}. This is also the case for N66A, which is 
marked by a prominent absorption lane of local dust crossing the whole  
nebula. The two other known examples of HEBs in the SMC are N88A and N81, which were  
also observed with {\it HST} \citep[][]{MHM99a,MHM99b,MHM02}.

HEBs are usually located adjacent to ordinary giant \hii\ regions or seen lying 
across them. This implies that they form as a consequence of triggering by 
a previous generation of massive stars in the complex.  
Simple reasoning suggests that HEBs and their small exciting clusters 
are formed from the material remaining after a preceding massive-star formation event. 
More specifically, the apparent association of N66A with the compressed ionized front and the 
absorption lane, both centered on the NGC\,346 cluster, suggests that N66A is a 
secondary, younger generation of stars. This implies that the exciting star(s) of N66A have 
formed according to the sequential star-formation model \citep[][]{Elmegreen77,Whitworth94}. 
However, since low-mass PMS candidates have also been reported in this region, 
other induced star formation scenarios, such as shock waves impacting molecular 
cores \citep[][]{Vanhala98} and/or radiation-driven implosion of molecular cores 
\citep[][]{Bertoldi89,Kessel03,Miao06,Mookerjea09} are or may have been at work. 
Several outstanding examples of 
triggered massive star formation in the Galaxy were recently studied by 
\citet[][and references therein]{Deharveng09}.

As shown above (Sect. 3), a couple of exciting stars are discovered inside N66A by 
high-resolution {\it HST} ACS images and NTT spectroscopy. 
Other undetected stars may be enshrouded in the gas and dust of N66A. 
In this respect \citet[][]{Simon07} report two YSO candidates towards N66A 
(Fig.\,\ref{fig:HST_stars}) from their analysis of the Spitzer survey.  
Using automatic processing, they detect a total of 1645 sources in the 
whole N66 region, among which some 50 embedded YSO candidates. 
Their automatic detection is based on  
spectral energy distribution (SED) fitting to 5 data points, which represent  
the fluxes in the Spitzer bands (at 3.6, 4.5, 5.8, 8.0, and 24 $\mu$m) 
with  spatial resolutions from 1\frac.6 to 6\frac\ (Sect. 2.3.2). Although the presence of 
YSOs is fully consistent with the star formation activity in N66A, 
their verification requires a detailed inspection of the Spitzer images. The Spitzer 
images in those bands uncover a particularly luminous object 
lying at the position of the compact \hii\ region N66A. It is indeed 
one of the most luminous IR sources in the whole field 
(13000 \slum ) and has a PSF of \ab\,3\frac.6 in its elongated direction. 
The second candidate YSO, lying at \ab\,2\frac.4 (2 pix) 
from the first object, cannot be firmly detected in the Spitzer images. 
No concrete substructure indeed shows up in the images on this scale. 
In addition, the derived IRAC colors for N66A,  [3.6]-[4.5] = 0.40 mag, [5.8]-[8.0] = 1.59 mag, and  
[3.6]-[8.0] = 3.56 mag (Sect. 2.3.2) do not satisfy the first criterion 
for  the YSO-class membership stated by   \citet[][]{Simon07}.    
From the [5.8]-[8] versus [3.6]-[4.5] color diagram and based on model 
calculations \citep[][]{Whitney04}, N66A also does not appear to belong definitely to the class   
of protostars. The IRAC colors of N66A 
agree perfectly with those of other HEBs, as studied 
by \citet[][]{Charmandaris08}. We note that N66A is above all a very bright 
\hii\ region with strong nebular emission lines affected by heavy extinction from
local dust.  Therefore, detecting a YSO inside the
\hii\ region using IRAC colors seems hazardous unless the YSO is the dominant source
inside N66A, which is obviously not the case. 
Anyhow,  the correspondence of the color of this bright \hii\ region with those of a YSO 
would be a coincidence.

The suggestion by the present work of a relatively young age for the exciting star(s) of 
N66A disagrees with the finding of \citet[][]{Sabbi07} according to which all 
the subclusters of the N66 complex (except one) have the same age, i.e. 3\,$\pm$\,1 Myr.  
\citet[][]{Sabbi07}'s subcluster Sc 10 overlaps N66A, in which they count a 
total of 61 low-mass PMS objects in an area 1.6 pc in radius.  
To estimate the ages, they fit the color-magnitude diagrams 
of the subclusters ({\it HST} ACS images F555W and F814W) with Padua isochrones 
\citep[][]{Bertelli94}. They also use PMS isochrones \citep[][]{Siess00} 
to evaluate the ages of the various subclusters. They note that  
compared to other subclusters, Sc 10 appears redder, suggesting either that it is 
a few million years older or that it is coeval with the others but
affected by higher extinction. They maintain that the magnitudes and colors of
the PMS stars in Sc 10 appear too bright and
red to be compatible with a stellar population older than \ab\,4 Myr. They therefore 
attribute an age of 3\,$\pm$\,1 Myr to this subcluster, as for others, and 
do not consider a younger age. At the same time, they note that 
``Some of these associations (i.e., Sc 10 and 12) appear still embedded
in dust and fuzzy nebulosities and are probably sites of recent or
even still ongoing star formation.'' This means that they do not preclude the possibility 
of a younger age. Once again, one should be cautious when interpreting 
the color-magnitude diagrams  of objects lying in the face of a very 
bright \hii\ region, because the colors may be contaminated by nebular emission.

Our suggestion of a younger age for the exciting stars of 
N66A, however agrees perfectly with other studies of  this region. In particular,     
\citet[][]{Gouliermis08} argue that the entire N66 region may host younger 
star formation events induced from the east, where the SNR B0057-724 lies  
\citep[][]{Reid06,Naze02,Danforth03,Naze04}. There is indeed a large \hi\ hole 
there, which is offset from the central parts of N66, and interesting enough, 
N66A lies on the triggered 
formation arc suggested to be associated with a shock wave coming from the direction of the 
SNR  (their Fig. 1). An expanding \hii\ region  or a bubble blown by the winds of the 
massive progenitor of the SNR B0057-724 and possibly the W-R binary HD\,5980 
and the O7 Iaf+ star Sk\,80 \citep[][]{Walborn90} may be the stimulating agent. 
There are therefore stars of a range of ages in the N66 complex, HD\,5980 and Sk\,80 being older 
than the NGC\,346 cluster, which has not yet produced any WN stars.

\section{Concluding remarks}

We have used imaging and spectroscopy in the optical with the ESO NTT as well  
{\it HST} ACS and Spitzer archive data to study N66A. This compact \hii\ region 
is quite a distinctive object in N66 (NGC\,346), the pre-eminent starburst 
region of the SMC. We have presented a global view of the whole region and emphasized 
the importance of N66A. We derived a number of the 
physical characteristics of N66A, and for the first time using spectroscopy 
the spectral classification of the main exciting star of N66A. 
It is a dwarf massive star of type earlier than O8. 

We have argued that N66A is probably produced by a recent massive star formation in N66. 
Its exciting stars are most likely to have been triggered by the action of shocks 
caused by a previous generation of massive stars.  
Moreover, N66A belongs to a rare class of compact \hii\ regions in the MCs, called 
HEBs (High-Excitation Blobs).  Only two other HEBs 
have so far been detected in the MCs.

\begin{acknowledgements}
We wish to thank Prof. Vassilis Charmandaris, Dept. of Physics, Univ. of Crete, Greece, 
for discussions and help in the exploitation of Spitzer data. We are also grateful to 
Dr. Nolan R. Walborn, Space Telescope Institute, Baltimore, USA, for advices about 
the spectral classification of the massive stars. We are indebted to 
Dr. Fr\'ed\'eric Meynadier, GEPI, Paris Observatory, France, for a preliminary 
reduction of the NTT images. And finally we would like to thank an anonymous referee 
whose helpful comments considerably improved the final version.

\end{acknowledgements}

\end{document}